\documentclass[10pt, sigconf,bookmarksnumbered,unicode,colorlinks=true,allcolors=blue]{acmart}
\usepackage[english]{babel}
\usepackage{amsmath, amsfonts, bm}
\usepackage{graphicx}
\usepackage{multirow}
\usepackage{array}
\usepackage{enumitem}
\usepackage{makecell}
\usepackage{subcaption}
\usepackage{tabularx}
\usepackage{pifont}

\acmConference[ICCAD '26]{2026 IEEE/ACM International Conference on Computer-Aided Design}%
  {November 8--12, 2026}{San Jose, CA, USA}

\title{PolySim: Deterministic Polynomial Surrogates for Cross-Modal Retrieval on CiM}
\makeatletter
\renewcommand{\@authorfont}{\small}          
\renewcommand{\@affiliationfont}{\normalsize} 
\makeatother

\author{Xinzhao Li$^{1}$, Charles Power$^{1}$, Pengyu Ren$^{2}$, Jongun Won$^{2}$, Likai Pei$^{2}$, Yuting Hu$^{3}$, Jinjun Xiong$^{3}$, Alptekin Vardar$^{4}$, Ningyuan Cao$^{2}$, Xiaobo Sharon Hu$^{2}$, Thomas Kämpfe$^{4,5}$, Kai Ni$^{2}$, Ruiyang Qin$^{1}$ \\ $^{1}$Villanova University, $^{2}$University of Notre Dame, $^{3}$University at Buffalo, $^{4}$Fraunhofer IPMS, $^{5}$TU Braunschweig}
\renewcommand{\shortauthors}{Li et al.}

\begin{abstract}
Cross-modal retrieval on edge devices benefits from probabilistic embeddings that capture semantic uncertainty, but deploying them on compute-in-memory (CiM) hardware remains an open problem. The core difficulty is a sampling gap: probabilistic methods such as PCME rely on Monte Carlo sampling and nonlinear distance evaluation at inference, which are fundamentally incompatible with CiM crossbar arrays that support only deterministic, single-step matrix-vector multiplication. Few existing probabilistic retrieval methods can be executed on a conventional crossbar. To bridge this gap, we propose PolySim, a framework that reformulates probabilistic retrieval into a fully deterministic pipeline. PolySim approximates each Gaussian embedding dimension using low-order polynomial bases and computes similarity via a learnable order-bilinear kernel, eliminating stochastic sampling while preserving distributional information. In experiments on six benchmarks spanning video, image, and audio retrieval, PolySim improves R@1 over deterministic baselines by up to 10.3\% and matches or exceeds PCME, while reducing inference to a single crossbar-compatible matrix-vector multiplication. CrossSim evaluation under realistic device non-idealities confirms robust deployment on conventional crossbar arrays. To the best of our knowledge, PolySim is the first method to enable probabilistic cross-modal retrieval on CiM hardware.

\end{abstract}

\begin{document}
\raggedbottom
\maketitle

\section{Introduction}
In the era of large-scale AI, customizing and personalizing pretrained AI models on edge devices is increasingly needed in areas such as healthcare, daily assistance, and companionship~\cite{qin2024enabling}. Due to the limited computational and storage resources on edge devices, retrieval-based approaches such as retrieval-augmented generation (RAG)~\cite{lewis2021retrievalaugmentedgenerationknowledgeintensivenlp} are better suited for on-device personalized AI than parameter tuning~\cite{qin2025empirical}. A major bottleneck of retrieval approaches on edge devices is the data movement latency between SSD and DRAM, which can even exceed the latency of parameter tuning itself. While this bottleneck has been addressed by hardware-software co-design under compute-in-memory (CiM) architectures~\cite{Xia2019,Jung2022}, where the information to be retrieved is shaped and stored on crossbar arrays~\cite{qin2024robust, qin2024nvcimptnvcimassistedprompttuning, li2026cq}, such co-design solutions have so far only been demonstrated for text-based retrieval tasks.

Under this trend, retrieval tasks are increasingly moving from text-only settings to cross-modal settings, where a text query must be matched against candidates from other modalities, such as images or audio. A central challenge in cross-modal retrieval is semantic uncertainty: a single video can be accurately described by many different sentences, and a single sentence may correspond to multiple valid video segments. To address this, the state-of-the-art approach employs probabilistic cross-modal embedding (PCME)~\cite{Chun_2021_CVPR, chun2024pcmepp}, which represents each sample as a Gaussian distribution rather than a single point, allowing variance to encode semantic ambiguity explicitly. PCME and its variants have consistently outperformed deterministic baselines across image--text, video--text, and audio--text benchmarks.

\begin{figure}[!t]  
    \centering
    \includegraphics[width=1.\linewidth]{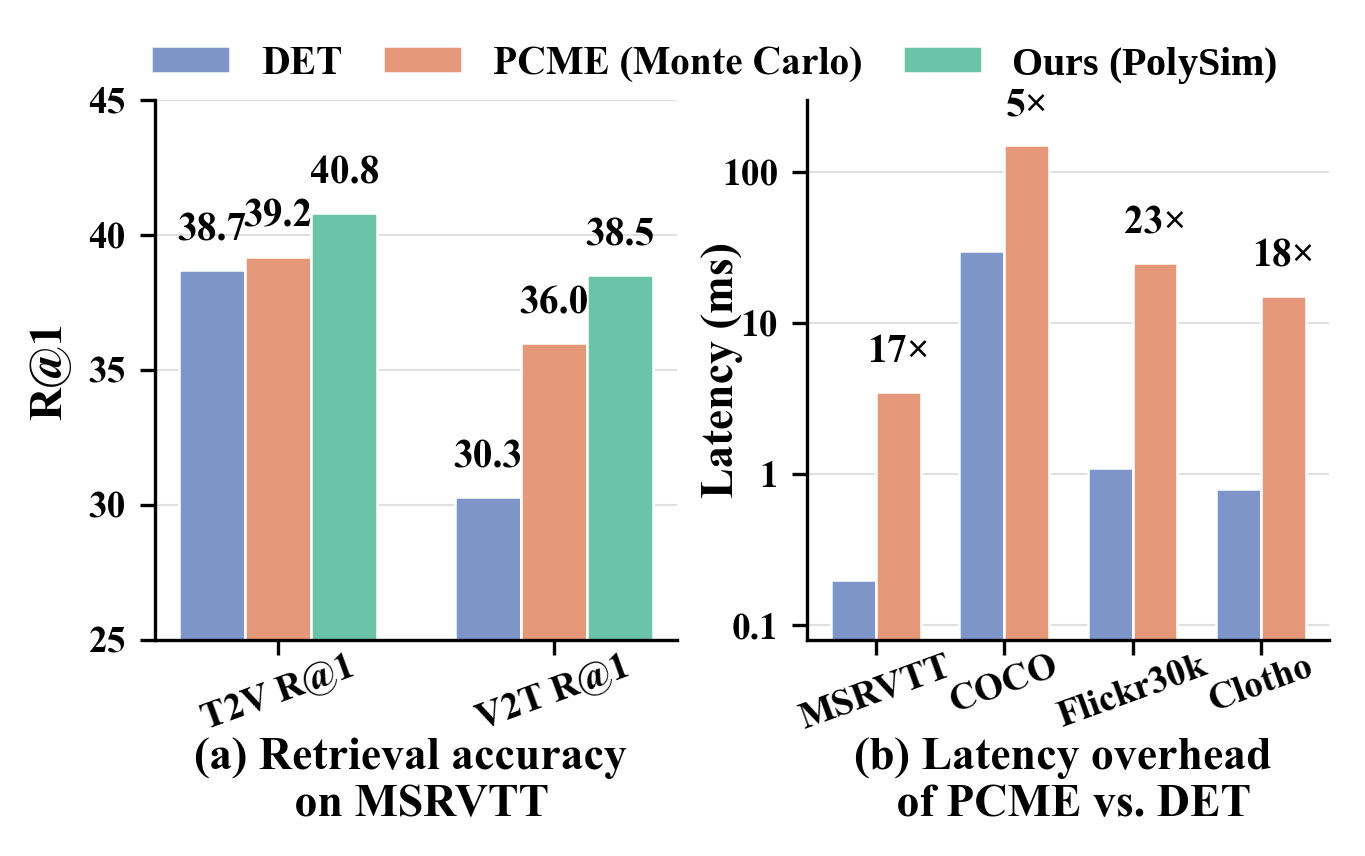}
    \vspace{-5ex}
    \caption{Probabilistic retrieval (PCME) improves accuracy over deterministic baselines but incurs 5–23× latency overhead due to Monte Carlo sampling. This sampling-based inference is fundamentally incompatible with CiM crossbar, motivating our deterministic reformulation.}
    \vspace{-3ex}
  \label{fig:profile_template}
  
\end{figure}

However, there are two major challenges in deploying PCME for on-device cross-modal retrieval. First, like text-based retrieval, PCME suffers from high data movement latency. Unlike deterministic embeddings, probabilistic embeddings are parameterized by both mean and variance and cannot be directly shaped into a single vector for crossbar storage. Second, even if the embeddings could be stored, the Monte Carlo sampling procedure inherent to PCME, requiring random number generation, $S^2$ pairwise distance evaluations per query--candidate pair, and nonlinear score accumulation, cannot be performed on standard crossbar arrays, which support only deterministic, single-step matrix--vector multiplication (MVM)~\cite{ren2026probabilistictreeinferenceenabled}. Supporting these operations would require additional off-chip components, negating the data movement advantage that motivates CiM in the first place. As shown in Figure~\ref{fig:profile_template}, PCME incurs 5--23× latency overhead over deterministic baselines across four benchmarks while providing moderate accuracy gains.

To bridge the gap between PCME and on-device cross-modal retrieval, we introduce \textbf{PolySim} (\textbf{Poly}nomial Surrogate \textbf{Sim}ilarity), a deterministic framework that enables probabilistic retrieval on standard crossbar arrays. Instead of maintaining the probabilistic embeddings and sampling from them at inference time, PolySim converts each per-dimension Gaussian density into a low-order Chebyshev polynomial expansion and concatenates the resulting coefficients into a deterministic surrogate embedding. Similarity is then computed using a learnable order-bilinear kernel that captures cross-order interactions among polynomial terms, and the entire inference path reduces to deterministic matrix multiplications that can be executed natively on a crossbar.


The contributions of this work can be summarized as follows:
\begin{itemize}
    \item To the best of our knowledge, this is the first work to enable probabilistic cross-modal retrieval on conventional CiM crossbar arrays.
    \item We propose PolySim, which converts Gaussian embeddings into deterministic polynomial coefficient vectors via Chebyshev expansion and computes similarity through a learnable order-bilinear kernel, reducing the entire inference to a single-step MVM on the crossbar.
    \item We validate PolySim on six benchmarks spanning video, image, and audio retrieval, achieving competitive or superior R@1 to PCME while maintaining robustness under realistic CiM device non-idealities.

\end{itemize}

\section{Background and Motivation}

\subsection{Cross-Modal Retrieval}

In text-based retrieval, each document is encoded into a fixed-size vector, and similarity with a query vector is computed via dot product. This dot product over the entire database reduces to a single matrix-vector multiplication (MVM), which maps directly onto CiM crossbar arrays. Prior work~\cite{qin2024robust, li2026cq} has demonstrated this path for text-based RAG acceleration on CiM.

Cross-modal retrieval extends this setting: given a text query, the system must find the most relevant candidate from a different modality, such as image, video, or audio~\cite{wang2025cross, qin2025tiny, luo2023open}. The key difficulty is semantic uncertainty. A single video can be described by many different sentences, and a single sentence may match multiple video segments. Standard point embeddings, where each sample maps to one vector, cannot capture this one-to-many structure.

To address this, PCME~\cite{Chun_2021_CVPR} represents each sample as a Gaussian distribution parameterized by $(\mu, \sigma)$, where the mean encodes semantic location and the variance encodes ambiguity. This improves retrieval accuracy, but changes the computational structure. The embedding is no longer a single vector, and computing similarity between two Gaussians requires Monte Carlo sampling rather than a single dot product. As a result, the simple MVM mapping that enables CiM acceleration for text-based retrieval no longer holds.

\subsection{Cost of Monte Carlo Sampling}
To compute similarity between two Gaussian embeddings, PCME draws $S$ samples from each distribution and averages over all pairwise distances:
\begin{equation}
\begin{aligned}
\mathrm{sim}(t,v)
&\approx
\frac{1}{S^2}
\sum_{i=1}^{S}
\sum_{j=1}^{S}
s\big(z_t^{(i)}, z_v^{(j)}\big), \\
z_t^{(i)} &\sim \mathcal{N}(\mu_t, \sigma_t),\;
z_v^{(j)} \sim \mathcal{N}(\mu_v, \sigma_v).
\end{aligned}
\end{equation}
For a database of $N$ entries with embedding dimension $D$, a single query incurs $O(N \cdot S^2 \cdot D)$ arithmetic and $O(N \cdot S \cdot D)$ data movement, compared with $O(N \cdot D)$ for a deterministic dot product. Moreover, $S$ must grow with the database size to maintain estimation accuracy, leading to super-linear scaling in both computation and data traffic.



\subsection{Crossbar-Based Compute-in-Memory}

As discussed in Section~2.1, text-based retrieval maps naturally onto CiM crossbar arrays because its core operation, a dot product between a query vector and a database matrix, is a single matrix-vector multiplication (MVM). A crossbar array performs this MVM in one cycle by encoding the matrix as NVM conductance values and applying the query as input voltages~\cite{Xia2019, Jung2022}.

However, crossbar arrays impose three constraints that are relevant to this work. First, the dataflow is deterministic and completes in a single step: the array cannot generate random numbers, iterate over multiple input vectors, or accumulate scores nonlinearly within a single operation. This is the fundamental reason Monte Carlo-based PCME inference cannot be executed directly on a crossbar. Second, NVM cells such as ReRAM and FeFET offer limited bit precision, typically 1--4 levels per cell, requiring the stored embeddings to be quantized. Third, device-to-device variations cause the programmed conductance to deviate from the target value, introducing noise into the MVM output. We evaluate the impact of the second and third constraints in the experiment sections.

\subsection{The Sampling Gap: Why Probabilistic Retrieval mismatches with Crossbar}

\begin{figure}[!t]
    \centering
    \includegraphics[width=.9\columnwidth]{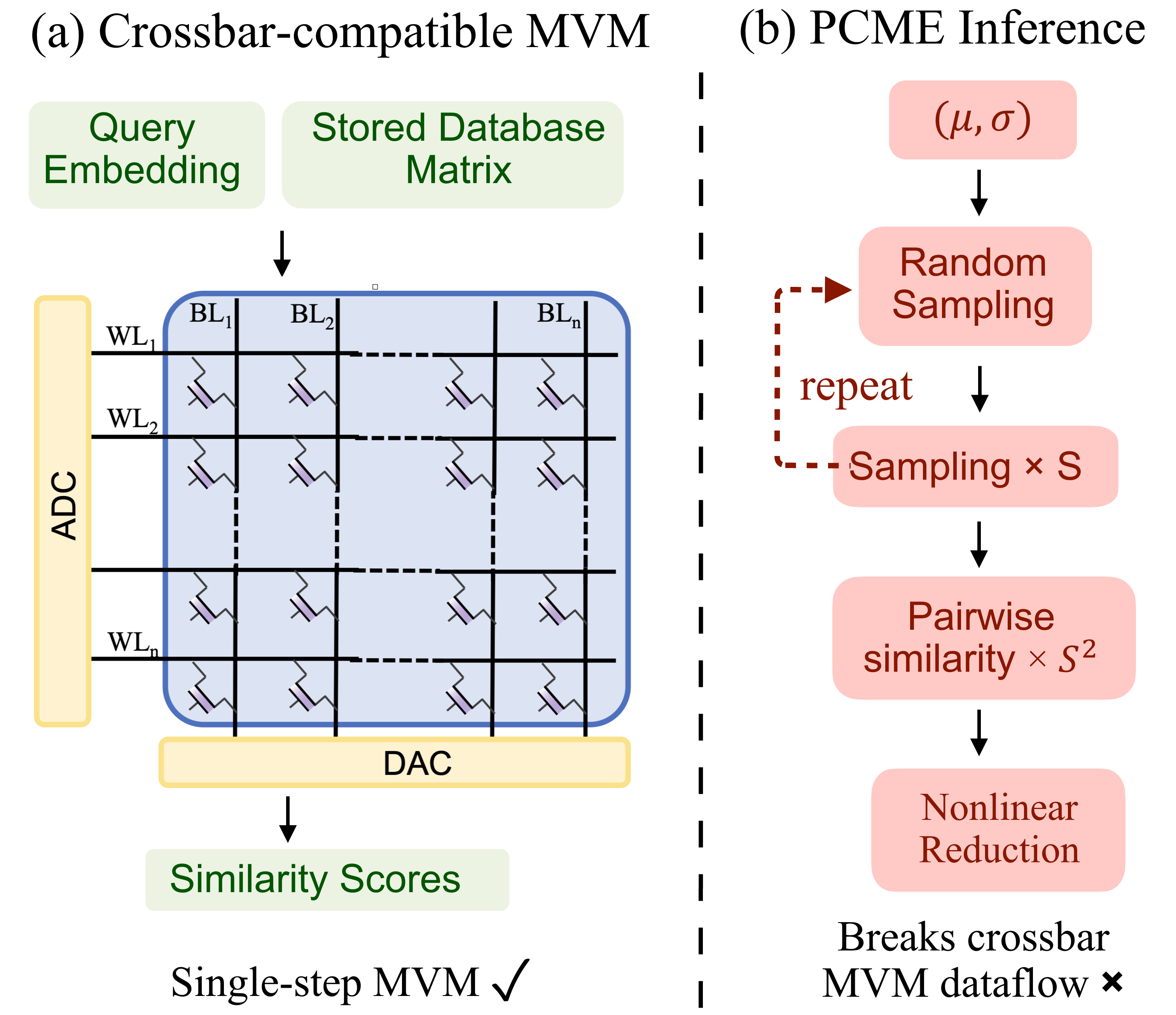}
    \vspace{-2ex}
    \caption{(a) Crossbar-compatible single-step MVM. (b) PCME requires off-chip sampling and nonlinear accumulation, breaking crossbar compatibility. Our method restores (a) while preserving distributional information.}
    \label{fig:sampling_gap}
\end{figure}

As illustrated in Figure~\ref{fig:sampling_gap}, every stage of Monte Carlo-based probabilistic inference conflicts with the crossbar operational model. The crossbar expects a single deterministic input vector per computation cycle and produces a linear output; it cannot generate random samples, cannot iterate over multiple sampled vectors within one pass, and cannot perform nonlinear reductions on the output currents~\cite{pei2026probabilistic}. Attempting to emulate Monte Carlo on a crossbar would require multiple separate passes per query--candidate pair, each preceded by off-chip sample generation and followed by off-chip nonlinear accumulation, completely negating the data movement advantage that motivates CiM in the first place. This is not a precision issue resolvable by improving NVM bit-density, nor a tiling issue addressable by scaling array dimensions. It is a structural mismatch between the stochastic, iterative, nonlinear nature of Monte Carlo inference and the deterministic, single-step, linear operation of a crossbar. Bridging this gap requires reformulating the retrieval algorithm itself so that the inference path is natively expressible as deterministic matrix multiplication.

\section{Proposed Work}

\begin{figure*}[t!]
\centering
\includegraphics[width=0.9\textwidth]{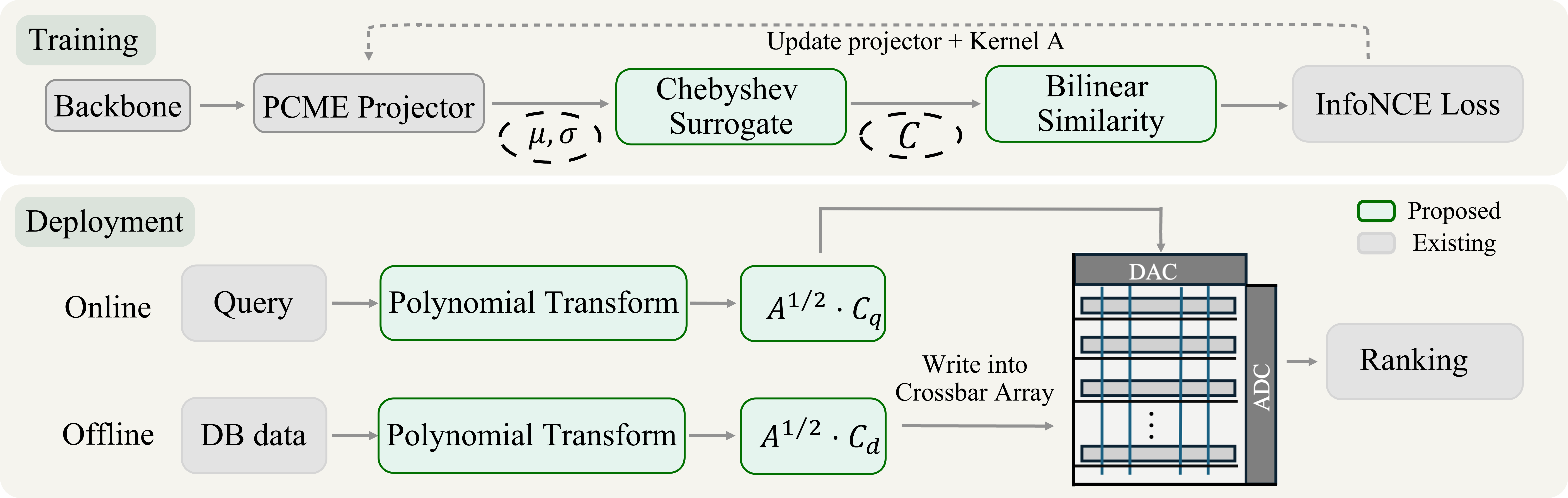}
\vspace{-2mm}
\caption{
Overview of the proposed framework. PolySim converts probabilistic Gaussian embeddings into a deterministic polynomial representation, enabling single-step MVM compatible with CiM hardware.
}
\label{fig:overview}
\vspace{-3mm}
\end{figure*}

\subsection{Overview}

This section presents the overall framework of PolySim, as illustrated in Figure \ref{fig:overview}. PolySim converts probabilistic Gaussian embeddings into a deterministic polynomial representation that enables matrix multiplication based similarity computation, while maximally preserving the structural properties of the original joint Gaussian embedding space.

The data flow passes through three main stages in the pipeline. First, the input multimodal data (e.g., text, image, or audio) is encoded by a backbone model, followed by a lightweight projection head that produces a Gaussian embedding parameterized by $(\mu, \sigma)$. Then, for each embedding dimension, the Gaussian density is approximated using a low-order Chebyshev expansion, and the resulting coefficients are concatenated to form a deterministic polynomial embedding, which preserves the structural characteristics of the original distribution while enabling efficient downstream processing.

To compute similarity, we operate directly on the polynomial embeddings using a learnable order-bilinear function implemented with matrix multiplications. During training, we adopt an asymmetric bidirectional InfoNCE~\cite{oord2019representationlearningcontrastivepredictive} objective to optimize the entire pipeline end-to-end, including the projection head and similarity parameters. During deployment, both query and database embeddings are deterministically transformed into polynomial coefficients, and similarity is computed using matrix operations, enabling efficient execution on hardware platforms such as CiM crossbar arrays.

\subsection{Gaussian Embedding Interface}

Our method takes as input a Gaussian embedding parameterized by a mean vector $\mu$ and a diagonal variance $\sigma$. In practice, such Gaussian embeddings are produced by appending a probabilistic projector, such as the PCME projector, to a multimodal backbone encoder.

In this work, we treat the Gaussian embedding $(\mu, \sigma)$ as the interface between probabilistic representation learning and deterministic inference. Rather than modifying the backbone encoder or the probabilistic projector itself, we operate on the Gaussian output of the projector and transform it into a deterministic polynomial representation for efficient similarity computation. This design allows PolySim to be attached after existing probabilistic projection modules while preserving their uncertainty-aware embedding space.

\subsection{Polynomial Surrogate Embedding}

We now introduce the core component of PolySim, which converts probabilistic Gaussian embeddings into deterministic polynomial representations. Given the Gaussian embedding $(\mu, \sigma)$, we construct a deterministic surrogate by approximating each per-dimension Gaussian density in a polynomial basis. For each embedding dimension $d$, we consider the one-dimensional density
\begin{equation}
g_d(x) = \frac{1}{\sqrt{2\pi}\sigma_d} \exp\left(-\frac{(x-\mu_d)^2}{2\sigma_d^2}\right),
\end{equation}
and approximate it using a truncated Chebyshev expansion.

Rather than computing coefficients through discrete projection, we adopt the Chebyshev recurrence formulation, which enables efficient and differentiable computation. The polynomial basis functions are generated via
\begin{equation}
\begin{aligned}
T_0(x) &= 1, \quad T_1(x) = x, \\
T_k(x) &= 2xT_{k-1}(x) - T_{k-2}(x).
\end{aligned}
\end{equation}
and the coefficients are obtained by evaluating the Gaussian density under this basis.

For each dimension, the resulting coefficients $\{c_{d,k}\}_{k=0}^{K}$ capture the shape of the underlying Gaussian distribution. These coefficients are concatenated across all dimensions to form the deterministic surrogate embedding,
\begin{equation}
C = [c_{1,0}, \dots, c_{1,K},\; \dots,\; c_{D,0}, \dots, c_{D,K}].
\end{equation}

This representation preserves distributional information in a structured and compact form. Moreover, low-order truncation ($K \ll D$) is sufficient in practice due to the approximation efficiency of Chebyshev polynomials, enabling a favorable trade-off between accuracy and computational cost.

\subsection{Order-Bilinear Similarity Function}

Given the deterministic polynomial embeddings, we define similarity directly in the coefficient space. A straightforward approach is to apply a dot product over the concatenated coefficients, which treats different polynomial orders independently and ignores their interactions. To address this limitation, we introduce a learnable order-bilinear similarity that models cross-order correlations through a parameter matrix.

For each embedding dimension $d$, let $c_t^{(d)}$ and $c_v^{(d)} \in \mathbb{R}^{K}$ denote the polynomial coefficients. The similarity is defined as
\begin{equation}
\mathrm{sim}(t,v)
=
\sum_{d=1}^{D}
\left(c_t^{(d)}\right)^\top A\, c_v^{(d)}
\;+\;
\gamma \left\langle \hat{\mu}_t,\hat{\mu}_v \right\rangle,
\end{equation}
where $A \in \mathbb{R}^{K \times K}$ is a learnable order-interaction matrix. In implementation, $A$ is initialized as either an identity matrix or a uniform matrix, and is symmetrized during training as $A \leftarrow \frac{1}{2}(A + A^\top)$ to ensure stable behavior. The second term is an optional residual similarity based on normalized mean embeddings, with a learnable weight $\gamma$.

Although the similarity is expressed in a bilinear form, it can be equivalently rewritten as a dot product in a transformed space. Specifically, by defining transformed embeddings as $\tilde{c} = A^{1/2} c$, the similarity becomes a standard inner product between $\tilde{c}_t$ and $\tilde{c}_v$. 

In practice, this transformation can be pre-applied to database embeddings offline, while query embeddings are transformed online. As a result, the final similarity computation reduces to a single matrix-vector multiplication, preserving compatibility with single-step CiM inference.

PolySim is trained end-to-end using an asymmetric bidirectional multi-positive InfoNCE objective. Given a batch of size $N$, we construct a similarity matrix $S \in \mathbb{R}^{N\times N}$. For one retrieval direction, the loss is defined as
\begin{equation}
\mathcal{L}_{\mathrm{mp}}(S)
=
-\frac{1}{N}\sum_{i=1}^{N}
\left[
\log \sum_{j \in \mathcal{P}(i)} e^{S_{ij}/\tau}
-
\log \sum_{j=1}^{N} e^{S_{ij}/\tau}
\right].
\end{equation}
where $\mathcal{P}(i)$ denotes samples sharing the same video ID as sample $i$. The final objective is
\begin{equation}
\mathcal{L}
=
\alpha\,\mathcal{L}_{\mathrm{mp}}(S)
+
\beta\,\mathcal{L}_{\mathrm{mp}}(S^\top).
\end{equation}

Gradients from the retrieval objective are backpropagated through the similarity function, directly updating the kernel matrix $A$, the polynomial coefficients, and the Gaussian projector. This enables joint optimization of representation and similarity under a fully deterministic pipeline.
\begin{figure}[!t]
\centering
\includegraphics[width=0.48\textwidth]{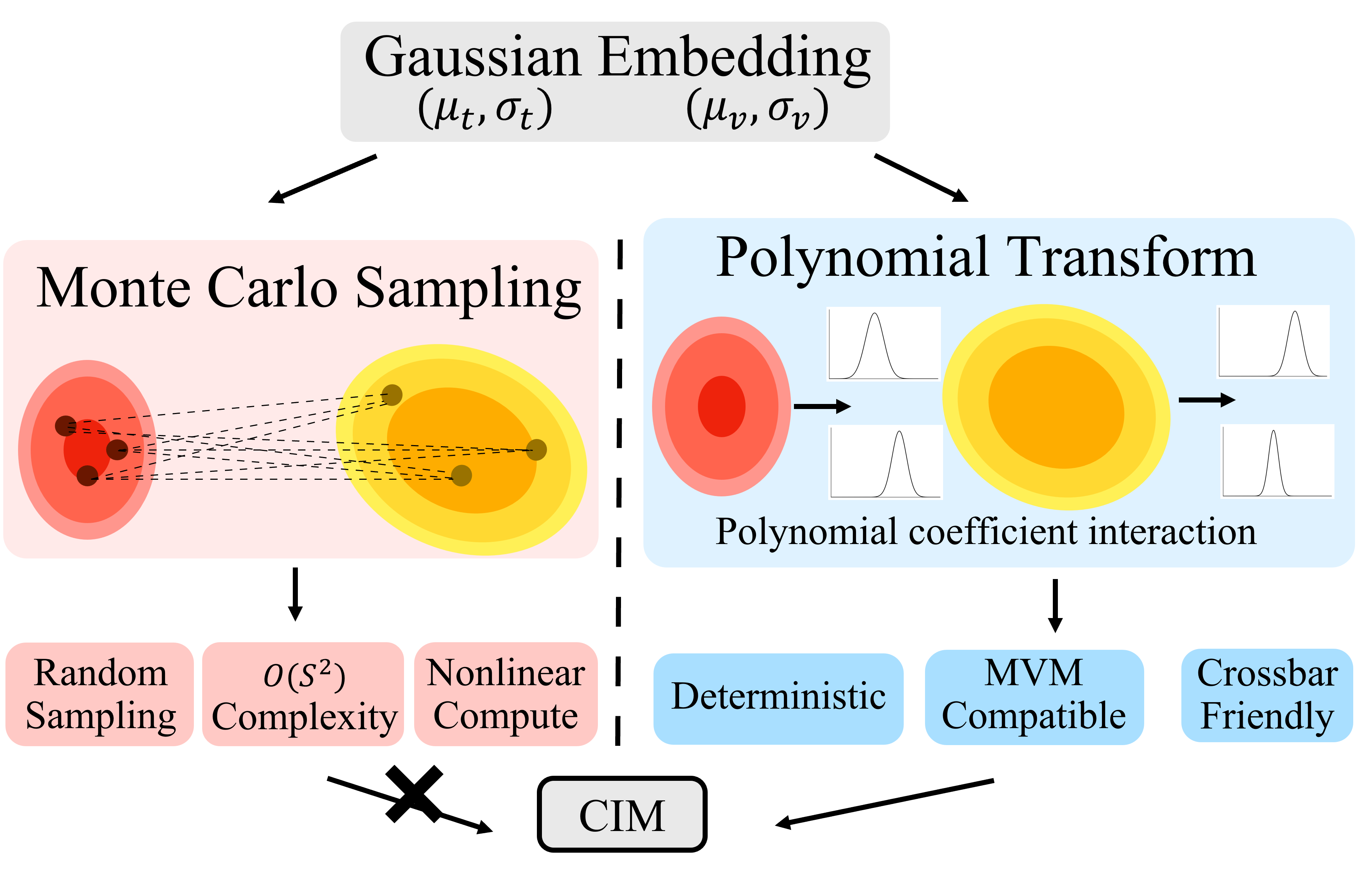}
\vspace{-5ex}
\caption{
Comparison of similarity computation. PCME relies on Monte Carlo Sampling. Our method instead uses polynomial coefficient.
}
\label{fig:similarity}
\vspace{-3mm}
\end{figure}

\section{Experimental Evaluation}
In this section, we present experiments to evaluate the performance, design choices, generalizability, and hardware efficiency of PolySim. Our primary task is multimodal retrieval, including video-to-text, image-to-text and audio-to-text retrieval. We first present the main results by comparing our method with DET and PCME across multiple benchmark datasets and base models. Then, we evaluate the hardware efficiency of the proposed method, enabled by its deterministic, matrix-multiplication-based inference. Finally, we conduct ablation studies to analyze key design components, including the choice of polynomial basis, polynomial order, and similarity function, providing further insights into the effectiveness of our design.

\subsection{Performance Evaluation}

To validate the effectiveness of PolySim, we evaluate it on six multimodal retrieval benchmarks: Clotho~\cite{drossos2019clothoaudiocaptioningdataset}, MSRVTT~\cite{xu2016msrvtt}, VATEX~\cite{wang2020vatexlargescalehighqualitymultilingual}, COCO~\cite{lin2015microsoftcococommonobjects}, Flickr30k~\cite{plummer2016flickr30kentitiescollectingregiontophrase}, and AudioCaps~\cite{kim-etal-2019-audiocaps}, covering video, image, and audio retrieval tasks. We consider multiple backbone models for each modality. For video and image retrieval, we use ImageBind~\cite{girdhar2023imagebind}, CLIP~\cite{radford2021clip}, and OpenCLIP~\cite{wu2023clap}, while for audio retrieval, we adopt CLAP-HTSAT and CLAP-Large. Since deploying probabilistic cross-modal retrieval on CiM has not been explored before, there is no directly comparable prior method. We adopt two natural reference points: deterministic embeddings (DET)~\cite{Chun_2021_CVPR}, which use the backbone's default point-vector output with dot-product similarity, and PCME~\cite{Chun_2021_CVPR}, which represents the state-of-the-art probabilistic approach. DET is CiM-compatible but discards distributional information; PCME preserves it but cannot be mapped to crossbar hardware.

\begin{table*}[!t]
\centering
\caption{
Retrieval performance of DET, PCME, and PolySim across video, image, and audio benchmarks. For each dataset, we report text-to-X (T2X) R@1, X-to-text (X2T) R@1 (\%), and software-side scoring latency$^\dagger$ (ms).
}
\vspace{-1.5ex}
\label{tab:main_results}
\fontsize{9pt}{8pt}\selectfont
\setlength{\tabcolsep}{2.5pt}
\renewcommand{\arraystretch}{1.15}
\begin{tabular*}{\textwidth}{@{\extracolsep{\fill}}
  ll
  rrr
  rrr
  rrr
  rrr
  rrr
  rrr
}
\toprule
& 
& \multicolumn{9}{c}{\textbf{MSRVTT}}
& \multicolumn{9}{c}{\textbf{VATEX}} \\
\cmidrule(lr){3-11} \cmidrule(lr){12-20}
& 
& \multicolumn{3}{c}{DET}
& \multicolumn{3}{c}{PCME}
& \multicolumn{3}{c}{PolySim}
& \multicolumn{3}{c}{DET}
& \multicolumn{3}{c}{PCME}
& \multicolumn{3}{c}{PolySim} \\
\cmidrule(lr){3-5} \cmidrule(lr){6-8} \cmidrule(lr){9-11}
\cmidrule(lr){12-14} \cmidrule(lr){15-17} \cmidrule(lr){18-20}
Task & Model
& T2V & V2T & Lat.$^\dagger$
& T2V & V2T & Lat.$^\dagger$
& T2V & V2T & Lat.$^\dagger$
& T2V & V2T & Lat.$^\dagger$
& T2V & V2T & Lat.$^\dagger$
& T2V & V2T & Lat.$^\dagger$ \\
\midrule
\multirow{3}{*}{Video}
& ImageBind
& 38.7 & 30.3 & 0.2 & 39.2 & 36.0 & 3.4 & \textbf{40.8} & \textbf{38.5} & 17.5
& 35.9 & 28.3 & 6.3 & 34.5 & 31.1 & 28.8 & \textbf{36.4} & \textbf{37.3} & 123.6 \\
& CLIP
& 30.5 & 26.4 & 0.2 & \textbf{31.6} & 29.3 & 1.9 & 30.3 & \textbf{31.4} & 7.6
& 22.9 & 19.3 & 5.5 & 22.6 & 21.0 & 31.1 & \textbf{24.6} & \textbf{23.9} & 62.4 \\
& OpenCLIP
& 34.8 & 28.0 & 0.2 & 34.3 & 35.6 & 1.9 & \textbf{35.3} & \textbf{36.3} & 7.5
& 24.6 & 18.6 & 5.6 & 21.9 & 20.2 & 31.1 & \textbf{25.8} & \textbf{25.8} & 62.5 \\
\midrule
& 
& \multicolumn{9}{c}{\textbf{COCO}}
& \multicolumn{9}{c}{\textbf{Flickr30k}} \\
\cmidrule(lr){3-11} \cmidrule(lr){12-20}
& 
& \multicolumn{3}{c}{DET}
& \multicolumn{3}{c}{PCME}
& \multicolumn{3}{c}{PolySim}
& \multicolumn{3}{c}{DET}
& \multicolumn{3}{c}{PCME}
& \multicolumn{3}{c}{PolySim} \\
\cmidrule(lr){3-5} \cmidrule(lr){6-8} \cmidrule(lr){9-11}
\cmidrule(lr){12-14} \cmidrule(lr){15-17} \cmidrule(lr){18-20}
Task & Model
& T2I & I2T & Lat.$^\dagger$
& T2I & I2T & Lat.$^\dagger$
& T2I & I2T & Lat.$^\dagger$
& T2I & I2T & Lat.$^\dagger$
& T2I & I2T & Lat.$^\dagger$
& T2I & I2T & Lat.$^\dagger$ \\
\midrule
\multirow{3}{*}{Image}
& ImageBind
& 50.4 & 66.8 & 32.1 & 51.8 & 66.1 & 151.0 & \textbf{54.4} & \textbf{69.8} & 963.9
& 78.1 & \textbf{92.6} & 1.1 & 78.6 & 92.4 & 25.2 & \textbf{80.2} & 92.2 & 67.0 \\
& CLIP
& 30.4 & 50.1 & 13.6 & 39.3 & 51.5 & 85.5 & \textbf{40.7} & \textbf{55.7} & 673.6
& 58.8 & 78.8 & 0.8 & 65.7 & 78.7 & 15.4 & \textbf{67.8} & \textbf{79.8} & 51.4 \\
& OpenCLIP
& 39.4 & 56.3 & 13.2 & 41.1 & 54.2 & 84.6 & \textbf{43.4} & \textbf{59.9} & 698.8
& 66.7 & 84.1 & 0.8 & 68.1 & 84.0 & 13.9 & \textbf{70.4} & \textbf{83.3} & 64.3 \\
\midrule
& 
& \multicolumn{9}{c}{\textbf{Clotho}}
& \multicolumn{9}{c}{\textbf{AudioCaps}} \\
\cmidrule(lr){3-11} \cmidrule(lr){12-20}
& 
& \multicolumn{3}{c}{DET}
& \multicolumn{3}{c}{PCME}
& \multicolumn{3}{c}{PolySim}
& \multicolumn{3}{c}{DET}
& \multicolumn{3}{c}{PCME}
& \multicolumn{3}{c}{PolySim} \\
\cmidrule(lr){3-5} \cmidrule(lr){6-8} \cmidrule(lr){9-11}
\cmidrule(lr){12-14} \cmidrule(lr){15-17} \cmidrule(lr){18-20}
Task & Model
& T2A & A2T & Lat.$^\dagger$
& T2A & A2T & Lat.$^\dagger$
& T2A & A2T & Lat.$^\dagger$
& T2A & A2T & Lat.$^\dagger$
& T2A & A2T & Lat.$^\dagger$
& T2A & A2T & Lat.$^\dagger$ \\
\midrule
\multirow{2}{*}{Audio}
& CLAP\_HTSAT
& 14.4 & 18.3 & 0.8 & \textbf{17.7} & \textbf{21.1} & 14.4 & 17.0 & 20.5 & 56.9
& 6.7 & 8.8 & 1.9 & 7.3 & \textbf{8.9} & 11.4 & \textbf{7.4} & 8.8 & 261.2 \\
& CLAP\_Large
& 14.3 & 21.3 & 0.8 & \textbf{18.2} & \textbf{22.9} & 14.7 & 17.9 & 21.7 & 55.6
& 5.6 & 8.5 & 1.9 & \textbf{7.4} & \textbf{9.3} & 11.2 & 7.2 & 9.1 & 215.0 \\
\bottomrule
\end{tabular*}
{\footnotesize $^\dagger$Latency is measured as software-side scoring cost in PyTorch. On a crossbar, PolySim reduces to 1 MVM per query (Table~\ref{tab:hw_comparison}), yielding $S^2\!\times$ lower hardware latency than PCME, where $S$ is the number of Monte Carlo samples.}
\vspace{-1.8ex}
\end{table*}

For each dataset, we report text-to-X (T2X), X-to-text (X2T), and software-side retrieval scoring latency. As shown in Table~\ref{tab:main_results}, PolySim consistently improves over DET across most datasets and models, with more pronounced gains in the X2T direction (e.g., video-to-text and audio-to-text), suggesting better preservation of distributional information. Compared with PCME, it achieves competitive or superior retrieval performance while avoiding stochastic sampling, and reformulates similarity computation into deterministic matrix-based operators that are compatible with CiM deployment.

Importantly, PolySim can outperform PCME even though both originate from the same probabilistic embedding formulation. Rather than approximating similarity through Monte Carlo estimation, our approach constructs a deterministic polynomial representation and learns the similarity directly in coefficient space, resulting in more stable and expressive similarity modeling.

The latency in Table~\ref{tab:main_results} reflects software-side scoring cost in the current PyTorch implementation (see table footnote). The higher software latency of PolySim is due to unoptimized polynomial expansion and bilinear evaluation on GPU. On CiM hardware, the relationship is reversed: PolySim requires only 1~MVM per query, whereas PCME requires $S^2$ passes (Table~\ref{tab:hw_comparison}), making PolySim faster by a factor of $S^2$ on the crossbar.

Table~\ref{tab:training_time} compares the per-epoch training time of PCME and PolySim. The Chebyshev-based approach incurs higher training cost due to polynomial expansion and order-bilinear modeling. However, this is a one-time overhead that enables fully deterministic single-step inference via matrix multiplication, making the resulting formulation more suitable for deployment.

Overall, these results demonstrate that our approach effectively combines the advantages of probabilistic embeddings with deterministic inference that is structurally aligned with CiM execution.

\subsection{CrossSim CIM Evaluation}

To evaluate the hardware deployability of PolySim, we conduct simulations using CrossSim~\cite{osti_code-73085} under realistic crossbar non-idealities. 
The level-dependent device variation profiles used in the conductance transition model are summarized in Table 2. The profiles for $RRAM_1$ and $RRAM_4$ are abstracted from measured RRAM devices~\cite{yao2020fully, liu2023architecture}, while $FeFET_2$ is based on measured FeFET data~\cite{wei2022switching}. $FeFET_3$ and $FeFET_6$ are extrapolated from these measured profiles.
While simulators such as NeuroSim~\cite{10.1109/TCAD.2020.3043731} can also evaluate inference accuracy, they primarily operate at the circuit and architecture level with higher-level abstractions. In contrast, CrossSim directly models analog matrix vector multiplication (MVM) distortions, which are the dominant source of error in our retrieval pipeline.

\newcolumntype{Y}{>{\centering\arraybackslash}X}

\begin{table}[!t]
\fontsize{9pt}{10pt}\selectfont
\centering
\caption{Level-dependent device variation profiles used in the conductance transition model. For multi-level cells, $L_0$--$L_3$ denote the nominal states and $\sigma_v$ their corresponding Gaussian deviations.}
\label{tab:var}
\vspace{-2ex}
\begin{tabularx}{\columnwidth}{c*{5}{Y}}
\toprule
\multirow{2}{*}{Name} & \multirow{2}{*}{\# of Levels} & \multicolumn{4}{c}{Device Variations $\sigma_v$} \\
& & $L_0$ & $L_1$ & $L_2$ & $L_3$ \\
\midrule
$RRAM_1$ (Device-1)  & 1 & 0.0100 & 0.0100 & 0.0100 & 0.0100 \\
$FeFET_2$ (Device-2) & 4 & 0.0067 & 0.0135 & 0.0135 & 0.0067 \\
$FeFET_3$ (Device-3) & 4 & 0.0049 & 0.0146 & 0.0146 & 0.0049 \\
$RRAM_4$ (Device-4)  & 4 & 0.0038 & 0.0151 & 0.0151 & 0.0038 \\
$FeFET_6$ (Device-5) & 4 & 0.0026 & 0.0155 & 0.0155 & 0.0026 \\
\bottomrule
\end{tabularx}
\end{table}


\begin{table}[!t]
\centering
\caption{Inference-time hardware requirements. $S$ denotes the number of Monte Carlo samples per distribution.}
\vspace{-1.5ex}
\label{tab:hw_comparison}
\fontsize{8pt}{9pt}\selectfont
\setlength{\tabcolsep}{5pt}
\renewcommand{\arraystretch}{1.15}
\begin{tabular}{lcccc}
\toprule
Method & \# MVM & RNG- & Linear & Crossbar \\
       & per query & free & accum. & compatible \\
\midrule
PCME   & $S^2$ & \ding{55} & \ding{55} & \ding{55} \\
PolySim   & 1     & \ding{51} & \ding{51} & \ding{51} \\
\bottomrule
\end{tabular}
\vspace{-1.5ex}
\end{table}

Table~\ref{tab:hw_comparison} compares the inference-time hardware requirements. Unlike PCME, which requires $S^2$ MVM passes, on-chip RNG, and nonlinear score reduction per query, PolySim reduces to a single MVM after offline database transformation and online query transformation, preserving standard single-step crossbar execution without any stochastic operations.

\begin{figure*}[!t]
\centering

\begin{subfigure}[t]{0.24\textwidth}
    \centering
    \includegraphics[width=\linewidth,
    trim=0 24 0 0,
    clip]{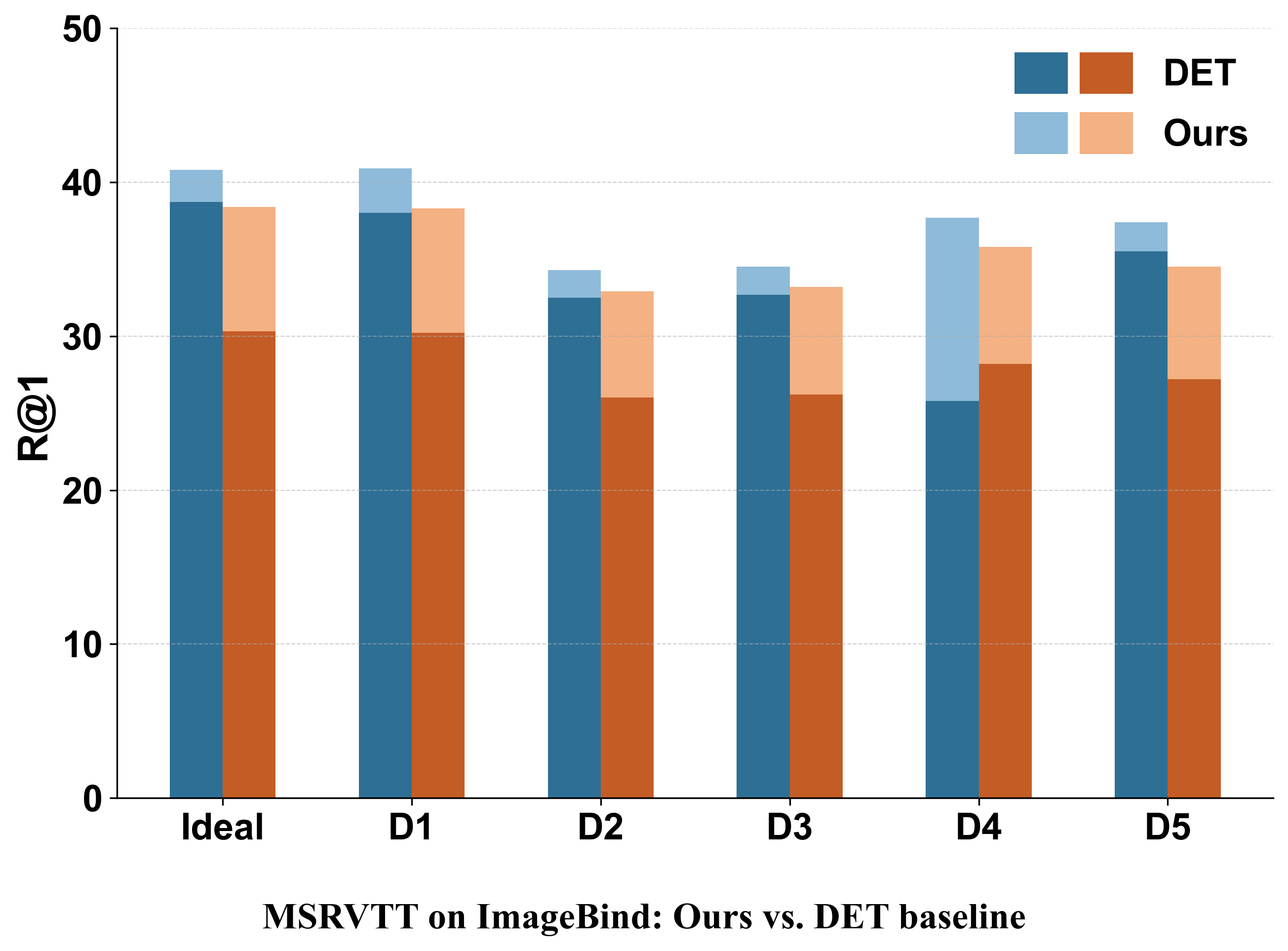}
    \vspace{-6mm}
    \caption{\scriptsize MSRVTT-Imagebind}
\end{subfigure}
\hfill
\begin{subfigure}[t]{0.24\textwidth}
    \centering
    \includegraphics[width=\linewidth,
    trim=0 24 0 0,
    clip]{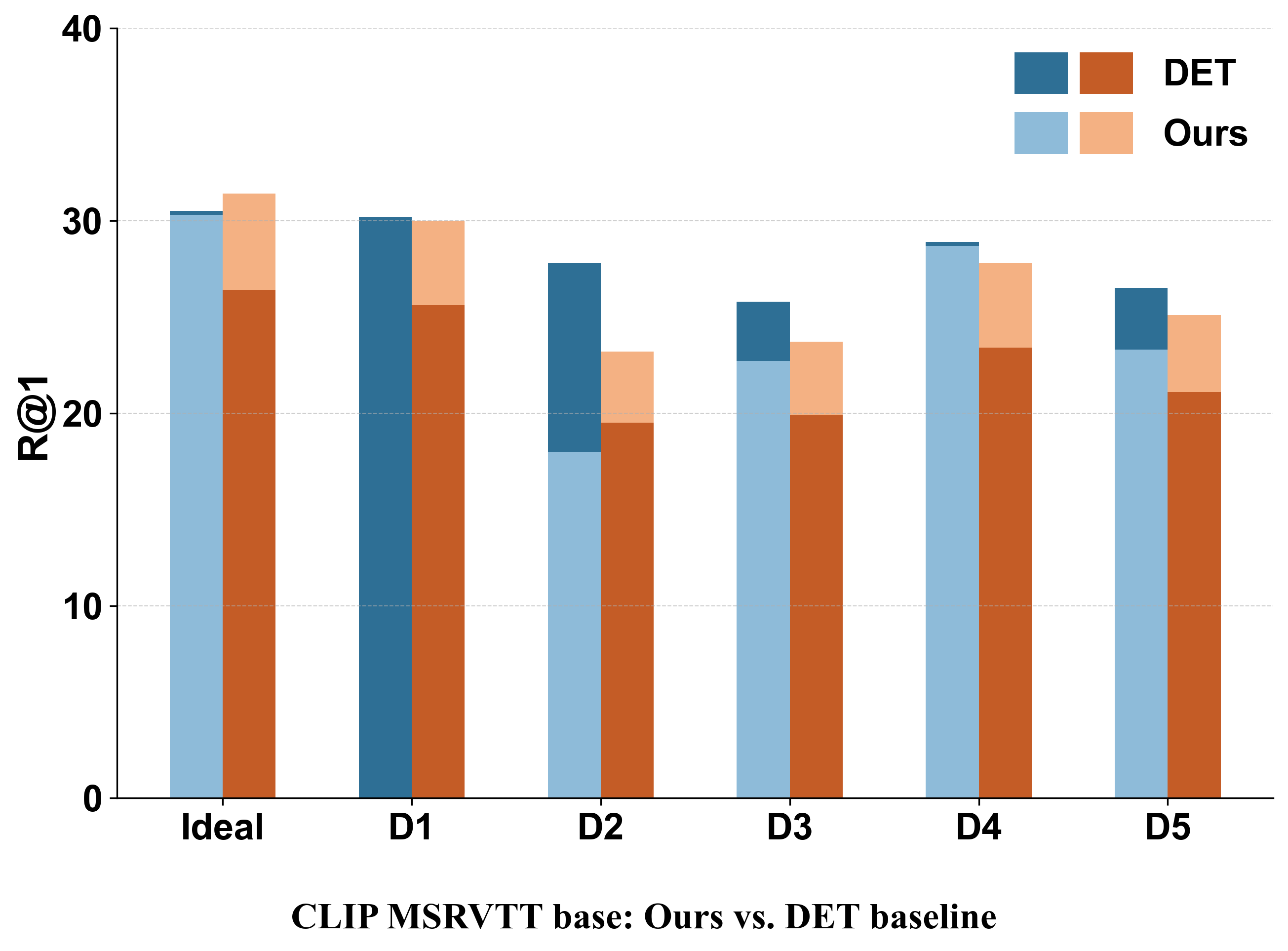}
    \vspace{-6mm}
    \caption{\scriptsize MSRVTT-CLIP}
\end{subfigure}
\hfill
\begin{subfigure}[t]{0.24\textwidth}
    \centering
    \includegraphics[width=\linewidth,
    trim=0 24 0 0,
    clip]{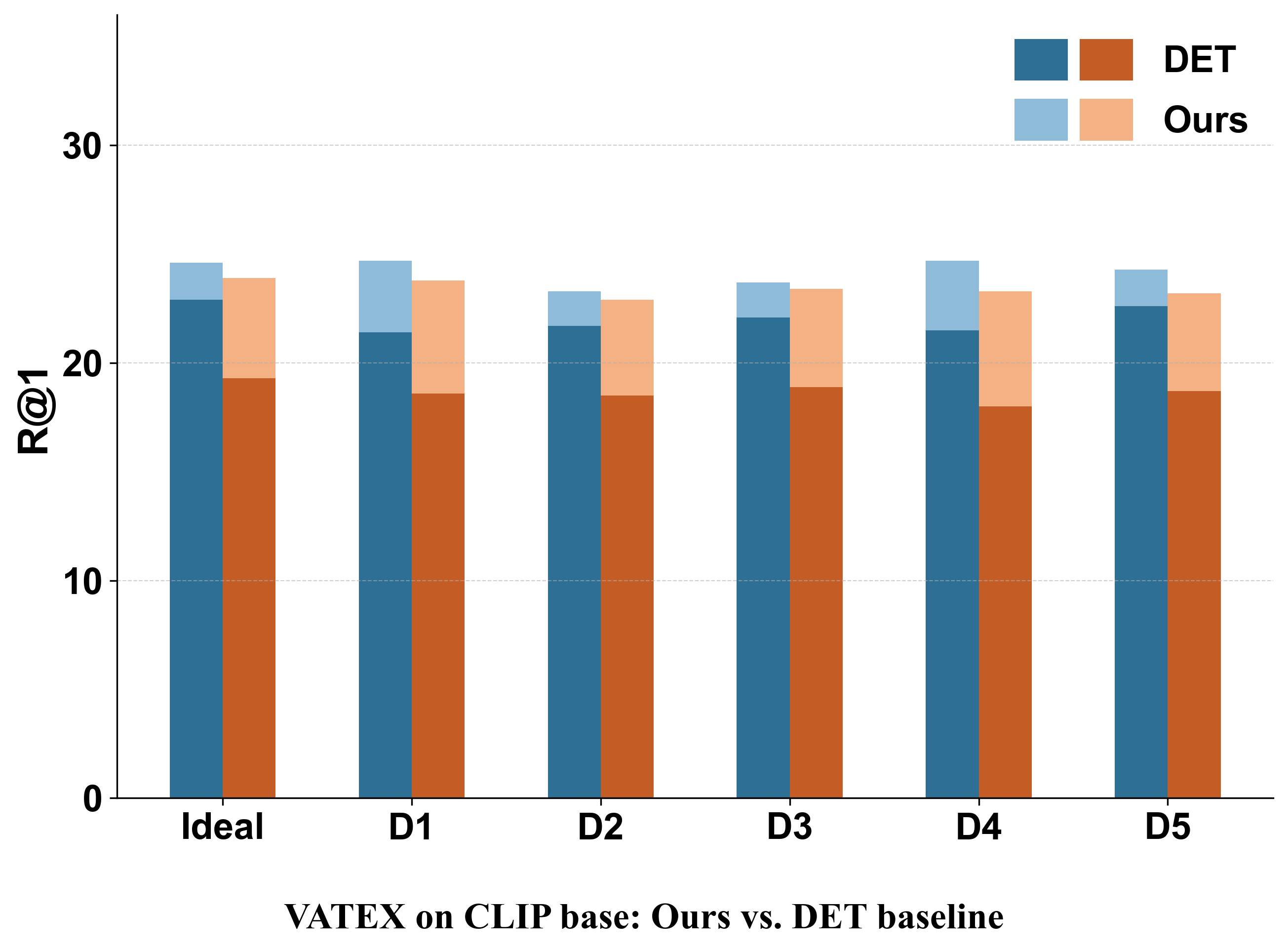}
    \vspace{-6mm}
    \caption{\scriptsize VATEX-CLIP}
\end{subfigure}
\hfill
\begin{subfigure}[t]{0.24\textwidth}
    \centering
    \includegraphics[width=\linewidth,
    trim=0 24 0 0,
    clip]{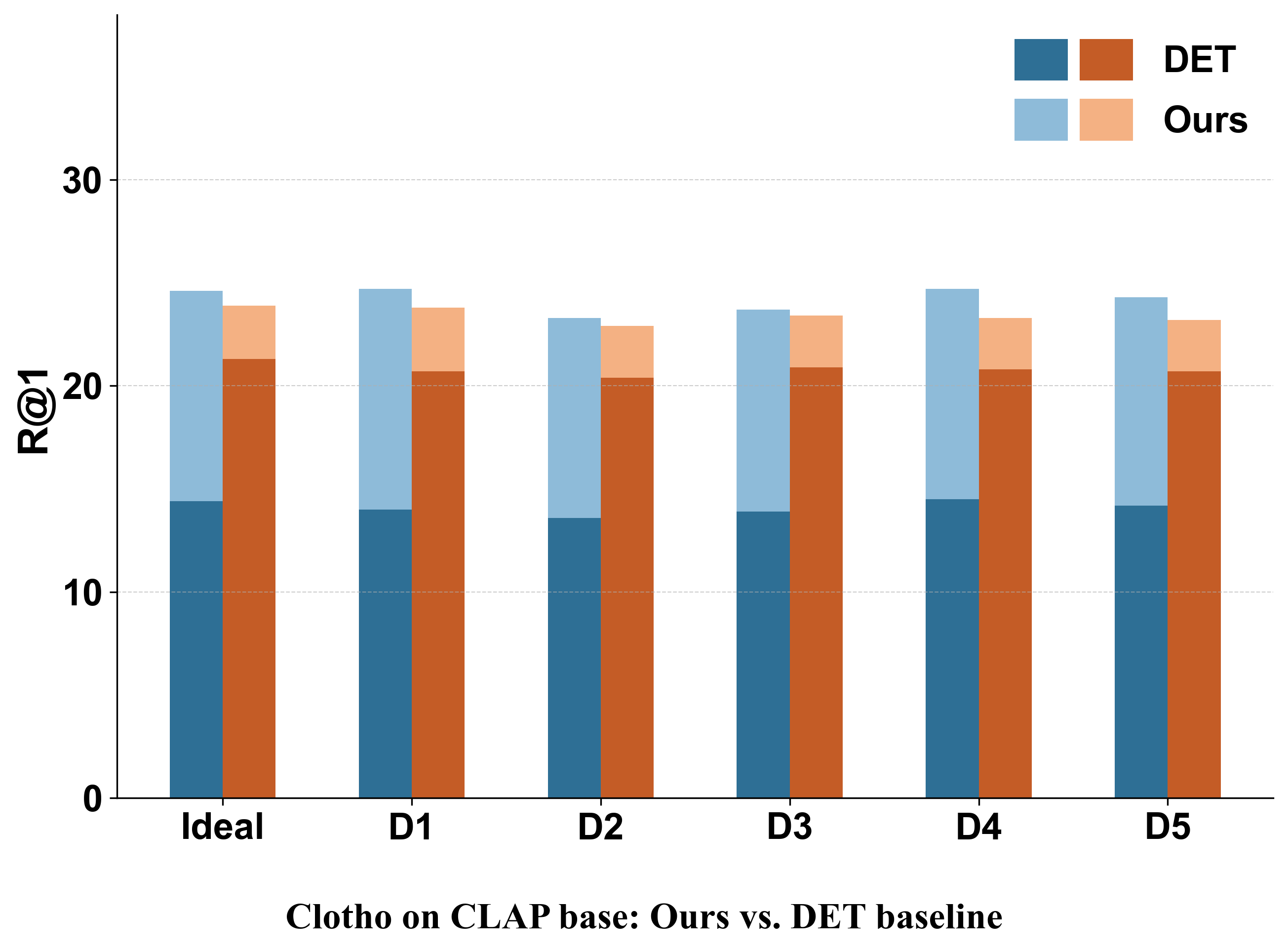}
    \vspace{-6mm}
    \caption{\scriptsize CLOTHO-CLAP}
\end{subfigure}

\vspace{-2mm}
\caption{
CrossSim evaluation under device non-idealities across different datasets and backbone models. Blue and orange bars denote the two retrieval directions for each dataset, while darker and lighter shades denote DET and Ours, respectively. D1--D5 denote the device variation configurations summarized in Table~\ref{tab:var}.
}
\label{fig:crosssim}
\vspace{-3mm}
\end{figure*}

Since our formulation reduces similarity computation to deterministic MVM operations, it can be directly mapped to crossbar-based CiM architectures and evaluated within CrossSim. In contrast, probabilistic approaches such as PCME rely on Monte Carlo sampling, which involves repeated random sampling and iterative similarity evaluation, and cannot be naturally expressed as a single-step MVM. As a result, they cannot be directly evaluated within the same simulation framework. Therefore, this experiment focuses on analyzing the robustness of PolySim under hardware non-idealities rather than comparing different algorithms.

We compare the retrieval performance between ideal GPU inference and CrossSim inference under multiple hardware configurations. As shown in Figure \ref{fig:crosssim}, PolySim maintains strong retrieval performance across different settings, with only moderate degradation under increasing non-idealities.

These results demonstrate that the proposed deterministic polynomial representation is not only effective in ideal settings but also robust under realistic hardware conditions, making it well-suited for deployment on CiM architectures.

\subsection{Similarity Function Design}


We next analyze the impact of the similarity function defined in Eq.~(4). While a standard dot product can be directly applied to the polynomial coefficient embedding, such a formulation treats different polynomial orders independently and ignores their structured interactions.

To evaluate the effectiveness of the proposed design, we compare three variants: (1) a naive dot product in the coefficient space, (2) a diagonal-weighted similarity that reweights each polynomial order independently, and (3) the proposed order-bilinear formulation. All variants share the same polynomial embedding and training setup, differing only in the similarity function.

As shown in Table~\ref{tab:similarity}, the naive dot product leads to noticeable performance degradation, indicating that the polynomial representation alone is insufficient to capture the underlying similarity structure. Introducing diagonal weighting significantly improves performance, demonstrating the importance of order-aware reweighting. However, it remains slightly inferior to the proposed formulation, suggesting that modeling interactions across polynomial orders provides additional benefits. In contrast, the proposed formulation consistently achieves the best performance, highlighting the advantage of capturing cross-order dependencies.

These results suggest that the performance gain does not solely come from the polynomial representation, but also critically depends on the similarity function design. By enabling interactions across different polynomial orders, the proposed formulation more accurately captures the structure of the underlying Gaussian embedding while remaining fully compatible with CiM inference.

\begin{table}[!t]
\centering
\caption{Ablation study of the similarity function.}
\vspace{-2ex}
\label{tab:similarity}

\setlength{\tabcolsep}{6pt}
\renewcommand{\arraystretch}{1.1}
\fontsize{8pt}{9pt}\selectfont
\begin{tabular}{l|cc|cc}
\toprule
\multirow{2}{*}{Method} & \multicolumn{2}{c|}{MSRVTT (ImageBind)} & \multicolumn{2}{c}{COCO (CLIP)} \\
& T2V & V2T & T2I & I2T \\
\midrule
Dot product & 0.1 & 0.1 & 1.5 & 1.8 \\
Diagonal matrix & 39.7 & 31.4 & 30.4 & 50.1 \\
\textbf{PolySim} & \textbf{40.8} & \textbf{38.5} & \textbf{40.7} & \textbf{55.7} \\
\bottomrule
\end{tabular}

\vspace{4mm}
\end{table}

\begin{figure}[!t]
\centering

\begin{subfigure}[t]{0.48\linewidth}
    \centering
    \includegraphics[width=\linewidth]{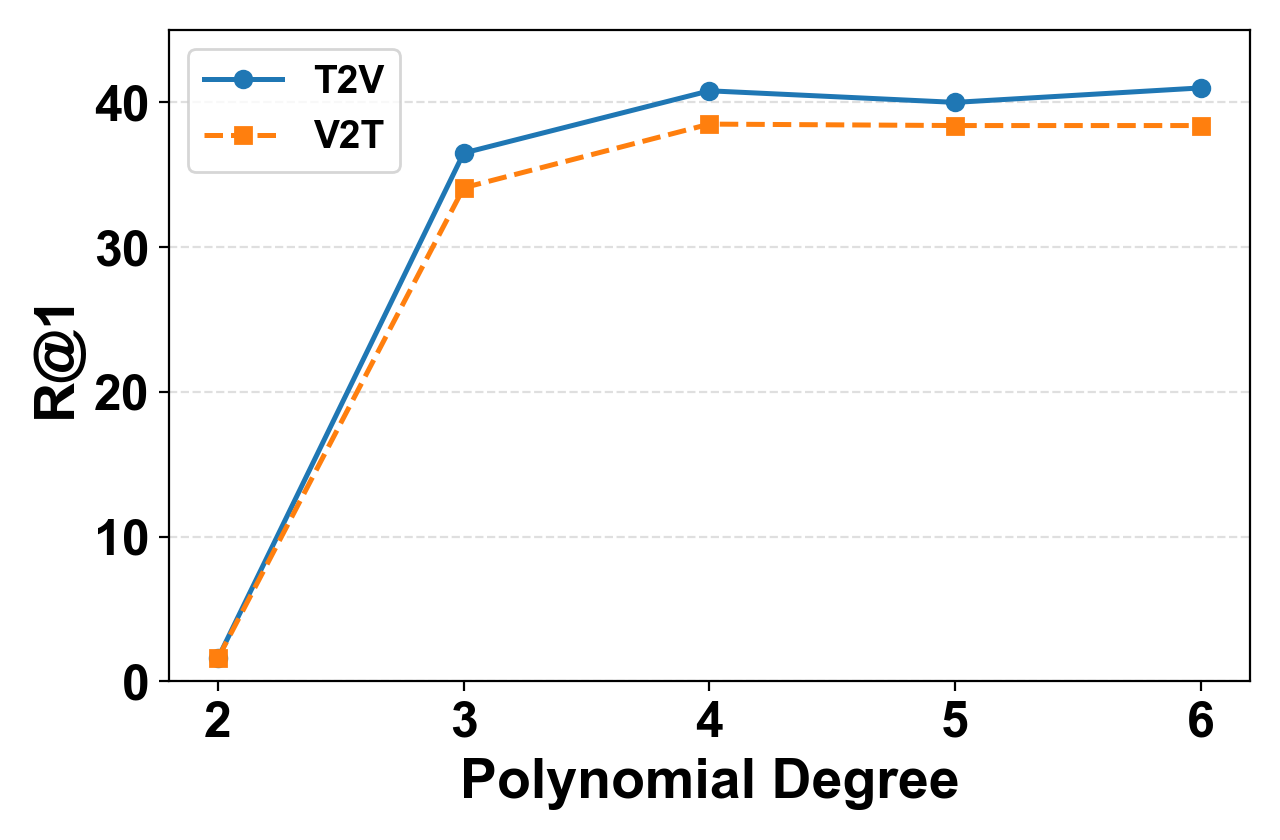}
    \vspace{-5mm}
    \caption{ImageBind on MSRVTT}
\end{subfigure}
\hfill
\begin{subfigure}[t]{0.48\linewidth}
    \centering
    \includegraphics[width=\linewidth]{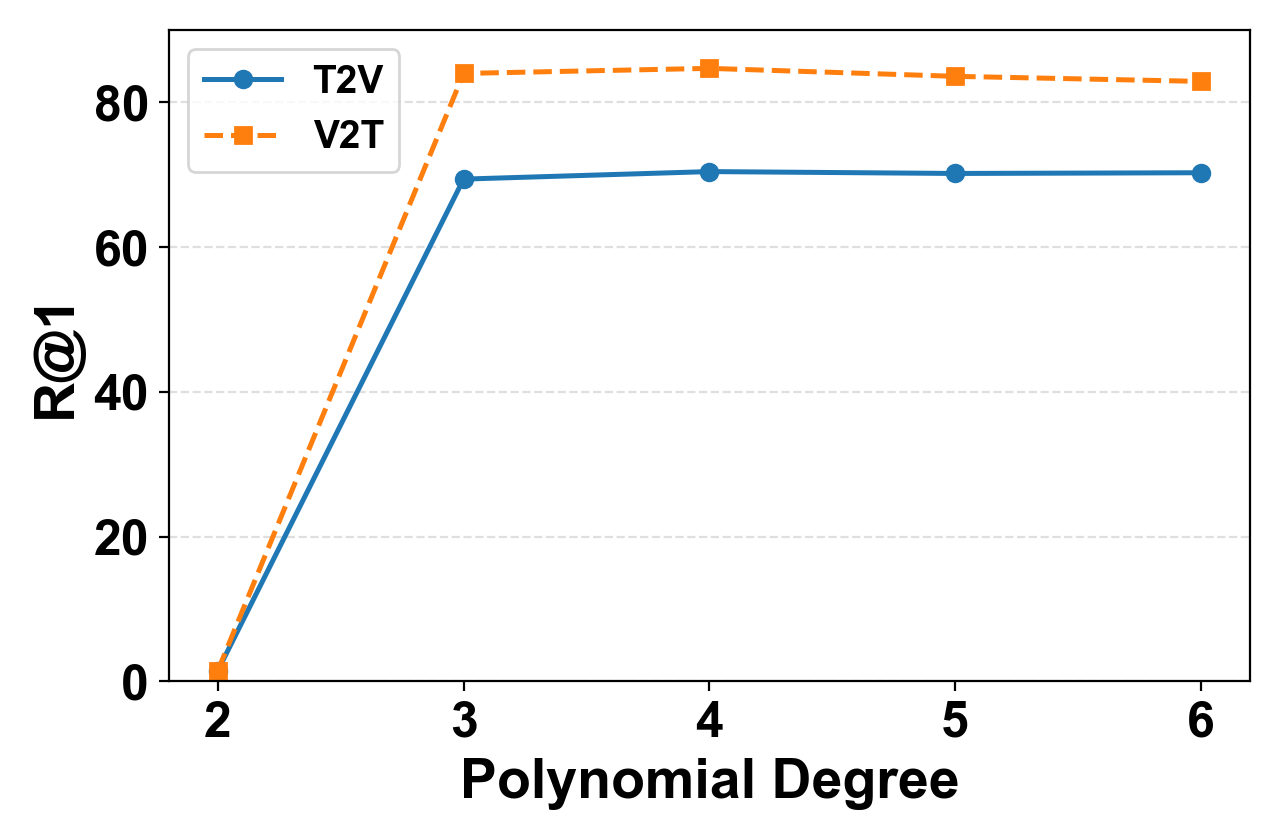}
    \vspace{-5mm}
    \caption{OpenCLIP on Flick}
\end{subfigure}

\vspace{-2mm}

\caption{
Effect of polynomial degree on retrieval performance (R@1) across different models and datasets.
}
\label{fig:degree}
\vspace{-3mm}
\end{figure}

\subsection{Polynomial Basis Design}

We analyze the design choices of the polynomial representation, including the approximation order and the choice of polynomial basis.

We first examine the impact of the polynomial degree on retrieval performance. As shown in Figure \ref{fig:degree}, increasing the degree from $K=2$ to $K=3$ leads to a significant improvement, indicating that very low-order expansions are insufficient to capture the underlying distribution. Beyond $K=3$, however, the performance quickly saturates, and further increasing the degree provides only marginal gains. This observation suggests that a compact low-order representation is sufficient in practice, which is desirable for efficient deployment.

We next compare different polynomial bases, including the raw power basis , defined as $x^k$, Legendre polynomials, and Chebyshev polynomials. For completeness, the Legendre polynomials are defined recursively as
\begin{equation}
\begin{aligned}
P_0(x) &= 1, \quad P_1(x) = x, \\
P_k(x) &= \frac{2k-1}{k}x P_{k-1}(x) - \frac{k-1}{k}P_{k-2}(x).
\end{aligned}
\end{equation}
Table~\ref{tab:basis} summarizes the results on both MSRVTT and VATEX datasets. Overall, all polynomial bases achieve comparable retrieval performance, indicating that the choice of basis has limited impact on accuracy. Nevertheless, Chebyshev consistently achieves competitive or slightly better performance across most metrics and datasets. In contrast, the raw power basis tends to be slightly less stable, while Legendre polynomials provide similar accuracy but without clear advantages.

Although the empirical performance differences are small, Chebyshev polynomials are theoretically well-suited for approximating smooth functions due to their minimax property, which minimizes the maximum approximation error over a bounded interval. Combined with the observed strong performance under low-order settings, this makes Chebyshev a more favorable choice for our framework. Importantly, this choice does not introduce additional computational overhead, while providing a more stable representation under low-order constraints.
\begin{table}[!t]
\centering
\fontsize{9pt}{9pt}\selectfont
\caption{
Comparison of polynomial bases on MSRVTT and VATEX datasets (R@1).
}
\label{tab:basis}
\vspace{-2mm}

\begin{tabular}{l|cc|cc}
\toprule
& \multicolumn{2}{c|}{MSRVTT (ImageBind)} 
& \multicolumn{2}{c}{VATEX (ImageBind)} \\
\cmidrule(lr){2-3} \cmidrule(lr){4-5}
Method 
& T2V@1 & V2T@1 
& T2V@1 & V2T@1 \\
\midrule

ImageBind 
& 38.7 & 30.3 
& 35.9 & 28.2 \\

Chebyshev 
& \textbf{40.8} & 38.5 
& \textbf{33.9} & 32.2 \\

Legendre 
& 40.0 & \textbf{38.6} 
& 33.4 & 32.1 \\

Raw 
& 39.5 & 38.0 
& 32.4 & \textbf{32.7} \\

\bottomrule
\end{tabular}

\vspace{1mm}
\end{table}

Overall, these results demonstrate that low-order polynomial expansion is sufficient for accurate representation, and that Chebyshev basis offers a robust and efficient design choice, making it particularly suitable for our deterministic and hardware-friendly formulation.

\subsection{Effect of Post-Training Quantization}

To evaluate the hardware deployment potential of PolySim, we analyze the effect of post-training quantization~\cite{banner2019posttraining4bitquantizationconvolution} (PTQ) on retrieval performance. In practical CiM systems, embeddings are typically represented with limited numerical precision. We therefore quantize the learned polynomial embeddings to different bit-widths and evaluate their performance without any retraining.

\begin{table}[!t]
\centering
\fontsize{8pt}{9pt}\selectfont
\caption{
Per-epoch training time (second) comparison between PolySim and PCME, measured on a single NVIDIA L4 GPU
}
\label{tab:training_time}
\vspace{-2mm}
\begin{tabular}{c c c c}
\toprule
\textbf{Dataset} & \textbf{Base Model} & \textbf{PCME} & \textbf{PolySim} \\
\midrule

\multirow{2}{*}{MSRVTT} 
& ImageBind & 5.04 & 10.05 \\
& CLIP      & 4.74 & 8.98 \\

\midrule

\multirow{2}{*}{COCO} 
& ImageBind & 9.54 & 32.16 \\
& CLIP      & 7.17 & 25.02 \\

\midrule

\multirow{1}{*}{Clotho} 
& CLAP\_HTSAT & 5.83 & 11.72 \\

\bottomrule
\end{tabular}
\vspace{-2ex}
\end{table}

\begin{figure}[!t]
\centering

\begin{subfigure}[t]{0.48\linewidth}
    \centering
    \includegraphics[width=\linewidth]{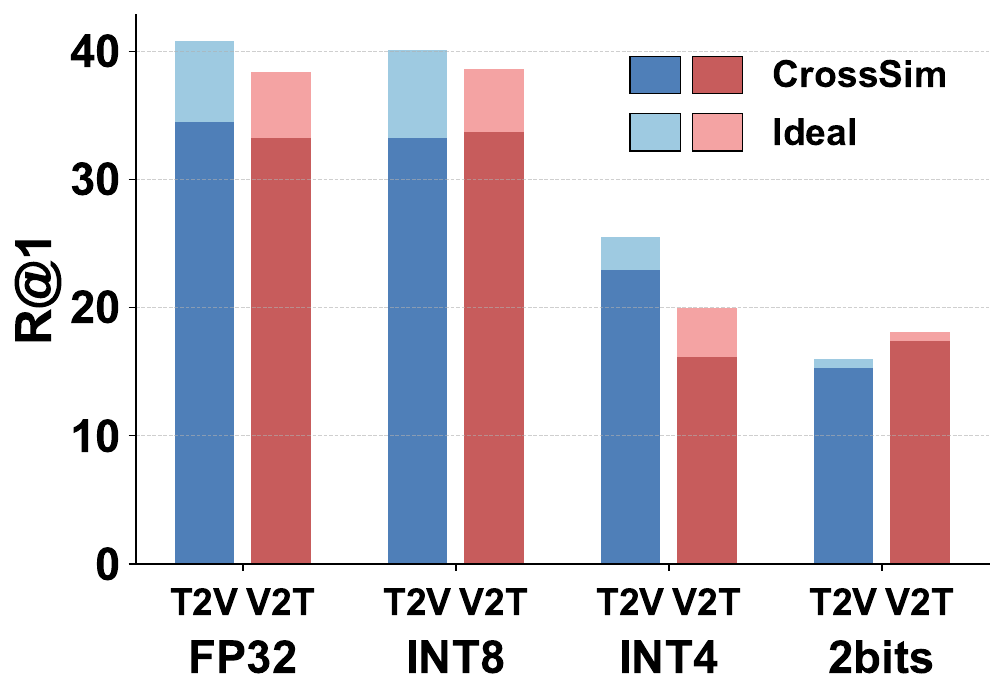}
    \vspace{-5mm}
    \caption{MSRVTT on ImageBind}
\end{subfigure}
\hfill
\begin{subfigure}[t]{0.48\linewidth}
    \centering
    \includegraphics[width=\linewidth]{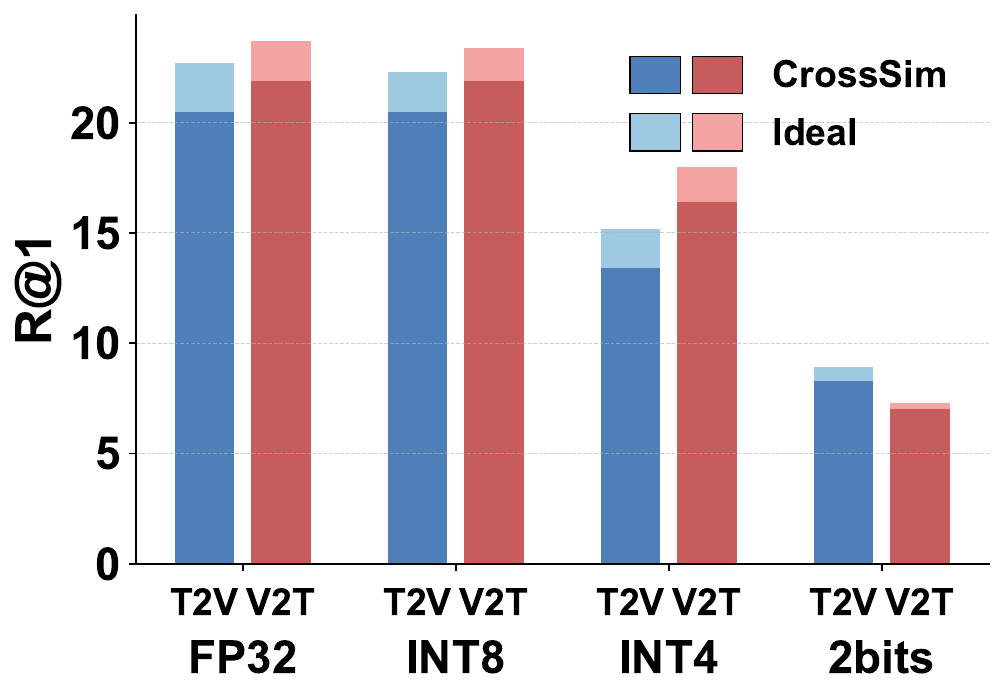}
    \vspace{-5mm}
    \caption{MSRVTT on CLIP}
\end{subfigure}

\vspace{-2mm}
\caption{R@1 retrieval under post-training quantization on MSRVTT.}
\label{fig:quant}
\vspace{-3mm}
\end{figure}

Figure~\ref{fig:quant} summarizes the results under varying quantization levels. PolySim remains robust under 8-bit quantization, with only negligible degradation relative to the full-precision setting, indicating that the learned embedding structure is well suited for moderate precision reduction. In contrast, more aggressive quantization, such as 4-bit and 2-bit, leads to substantial performance loss. This suggests that PolySim is suitable for moderate-precision deployment, but is not naturally optimized for ultra-low-bit regimes. In particular, aggressive quantization of the polynomial coefficients and the bilinear interaction matrix can significantly distort the learned similarity geometry, making 2-bit deployment challenging even with quantization-aware optimization. 

These results reveal the practical operating boundary of conventional crossbar arrays for cross-modal retrieval. At 8-bit precision, PolySim delivers near-lossless retrieval, confirming that standard multi-level NVM cells are sufficient for this task. The performance drop at 4-bit and 2-bit reflects a fundamental precision limit of conventional crossbar hardware, not a limitation of the PolySim formulation itself. For deployment scenarios where moderate precision is acceptable, conventional crossbar arrays offer a compelling combination of maturity, manufacturability, and energy efficiency. For scenarios demanding ultra-low-bit operation, these results motivate the co-design of novel device architectures, such as multi-bit analog cells or hybrid digital-analog arrays, tailored to the polynomial embedding structure introduced by PolySim.

\section{Conclusion}
We presented a deterministic surrogate framework for probabilistic cross-modal retrieval. By replacing Monte Carlo inference over Gaussian embeddings with low-order Chebyshev coefficients and a learnable order-bilinear similarity, PolySim preserves uncertainty-aware information, improves retrieval over deterministic baselines, and provides a more hardware-compatible inference path. This offers a practical bridge between probabilistic embedding quality and CiM-oriented deployment. More broadly, by establishing the first end-to-end pathway from probabilistic embeddings to crossbar execution, PolySim opens cross-modal retrieval as a new application domain for CiM architectures and provides a quantitative reference for the precision requirements of future CiM hardware design.

\newpage
\bibliographystyle{unsrt}
\bibliography{citations}

\end{document}